\documentclass[10pt, preprint]{aastex}

\usepackage{graphicx}
\usepackage{epsfig}

\newcommand{\br}{\mbox{{\boldmath$r$}}} 
\newcommand{\bs}{{\bf s}}

\newcommand{\cut}{ \omega_a }

\newcommand{\vst}{ v^*}
\newcommand{\pst}{ p^*}

\begin{document}

\title{IMPACT OF LOCALLY SUPPRESSED WAVE SOURCES ON HELIOSEISMIC TRAVEL TIMES}
\shorttitle{IMPACT OF LOCALLY SUPPRESSED WAVE SOURCES}
\author{S. M. Hanasoge, S. Couvidat, S. P. Rajaguru}
\affil{W. W. Hansen Experimental Physics Laboratory, Stanford University, Stanford, CA 94305}
\email{shravan@stanford.edu}
\and
\author{A. C. Birch}
\affil{Colorado Research Associates, NWRA, Colorado 80301}

\begin{abstract}
Wave travel-time shifts in the vicinity of sunspots are typically interpreted as arising predominantly from magnetic fields, flows, and local changes in sound speed. We show here that the suppression of granulation related wave sources in a sunspot can also contribute significantly to these travel-time shifts, and in some cases, an asymmetry between in and outgoing wave travel times. The tight connection between the physical interpretation of travel times and source-distribution homogeneity is confirmed. Statistically significant travel-time shifts are recovered upon numerically simulating wave propagation in the presence of a localized decrease in source strength. We also demonstrate that these time shifts are relatively sensitive to the modal damping rates; thus we are only able to place bounds on the magnitude of this effect. We see a systematic reduction of 10-15 seconds in $p$-mode mean travel times at short distances ($\sim 6.2$ Mm) that could be misinterpreted as arising from a shallow (thickness of 1.5 Mm) increase ($\sim$ 4\%) in the sound speed. At larger travel distances ($\sim 24$ Mm) a 6-13 s difference between the ingoing and outgoing wave travel times is observed; this could mistakenly be interpreted as being caused by flows.


\end{abstract}
\section{INTRODUCTION}
The discovery that sunspots support oscillations \citep[e.g., see the review by][]{bogdan06} was important for it introduced the possibility of using measurements of phase-shifts in the propagating waves to probe the underlying structure and dynamics of these enigmatic objects. Some of our current observational understanding of the sunspot interior comes from inverse theory applied in conjunction with time-distance helioseismology \citep{duvall,gizon05} on waves in these regions. Subsequent to the studies of flows in and around sunspots by \citet{Duvall1996}, inversions utilizing the ray \citep{Kosovichev1997}, Rytov \citep{jensen03b}, and Born \citep{birch} approximations were performed to recover the interior structure of sunspots \citep{Kosovichev2000, jensen03a, couvidat}. In recent years, many of these results have come into question because the analyses do not explicitly account for the influence of magnetic fields on wave travel times. Apart from direct mechanical effects on the waves, magnetic fields are also responsible for impeding the action of near-surface convection, commonly believed to be the source of waves \citep[e.g.,][]{stein00}. Despite the work of \citet{woodard} and \citet{Gizon2002}, causal factors of travel-time shifts such as wave damping and source distribution inhomogeneity in the context of sunspots have not been studied in detail.

\citet{Gizon2002} first derived the linear sensitivity of $f$-mode travel times to local changes in source strength, later corroborated through time-distance analyses of artificial data by \citet{hanasoge07}. The concept of variations in source-strength engendering travel-time shifts can be somewhat mystifying. Surely waves do not speed up or slow down when a source emits a wave of half the amplitude, as the naive interpretation seems to indicate? The answer lies in the manner in which travel times are computed; stripped of physical interpretation, travel times are obtained by fitting cross correlations of velocity (or intensity) signals between pairs of points or a point and an annulus. The measurement points do not constitute a source-receiver pair as in the typical geophysical situation; rather, all waves that contain coherent phase information at these points contribute to the cross correlations. The wave travel times measured in a system with a spatially uniform distribution of sources and a specific set of damping rates have certain expectation values. However, it is conceivable that over a region where the directionality or spatial distribution of waves is biased, the contributions by wave packets (to the cross correlations) from disparate directions and points are not in the same proportion as in the spatially uniform case. Consequently, there is a shift in the expectation value of the travel time in this region. In fact, the term `travel time' is better interpreted as a quantity that describes the statistics of the wave field than the physical wave travel time between the measurement points. Damping also plays an important role, for it determines the extent of coherence of the waves and the degree of contribution to the cross correlations. These can be serious issues in sunspots, because of the possible lack of sources and the putative excesses in damping and absorption \citep[e.g.,][]{braun87}. 

Mean travel times are defined as the average of the in- and outgoing wave travel times, while difference travel times are obtained by subtracting the two. We posit that the classical interpretation of mean travel-time shifts as mostly arising from changes in the sound speed and difference travel-time shifts predominantly from flows in sunspots is incomplete because the lack of wave sources can also cause significant mean and difference travel-time shifts; this effect is demonstrated here via numerical simulations and semi-analytical methods. In $\S$\ref{forward.model}, we describe the numerical machinery employed to perform the simulations and discuss the impact of horizontal boundary conditions on the resultant time shifts. In order to characterize the influence of damping rates, we apply the semi-analytical techniques of \citet{Gizon2002}. We analyze the simulated data with methods of time-distance helioseismology in $\S$\ref{travel.times}; comparisons are drawn between the results of simulations and the semi-analytical models. Finally, we summarize and conclude in $\S$\ref{discussion}.


\section{NUMERICAL PROCEDURE AND TEST CASES\label{forward.model}}
Using techniques developed in \citet{hanasoge1}, \citet{hanasoge07}, and \citet{hanasoge_thesis}, wave propagation in the near-surface layers of the Sun is simulated in a box of dimension $400 \times 400 \times 35~ {\rm Mm}^3$, where the third dimension is depth. The background stratification is a convectively stabilized form of model S \citep{jcd}, described in Appendix~\ref{alt.model}. Waves are stochastically excited by introducing a forcing term in the vertical momentum equation; the forcing function is prescribed such that a solar-like power spectral distribution is obtained. The solution is temporally evolved using a second-order optimized Runge-Kutta integrator \citep{hu}. The vertical derivative is resolved using sixth-order compact finite differences with fifth-order accurate boundary conditions \citep{hurl}. Depending on the choice of boundary conditions, the derivatives in the horizontal directions are computed either using these compact finite differences (absorbing conditions) or the Fast Fourier Transform (periodic boundaries). The validation and verification of the code is discussed in Appendix~\ref{verify}.

\begin{figure}[!ht]
\centering
\epsscale{1.0}
\plotone{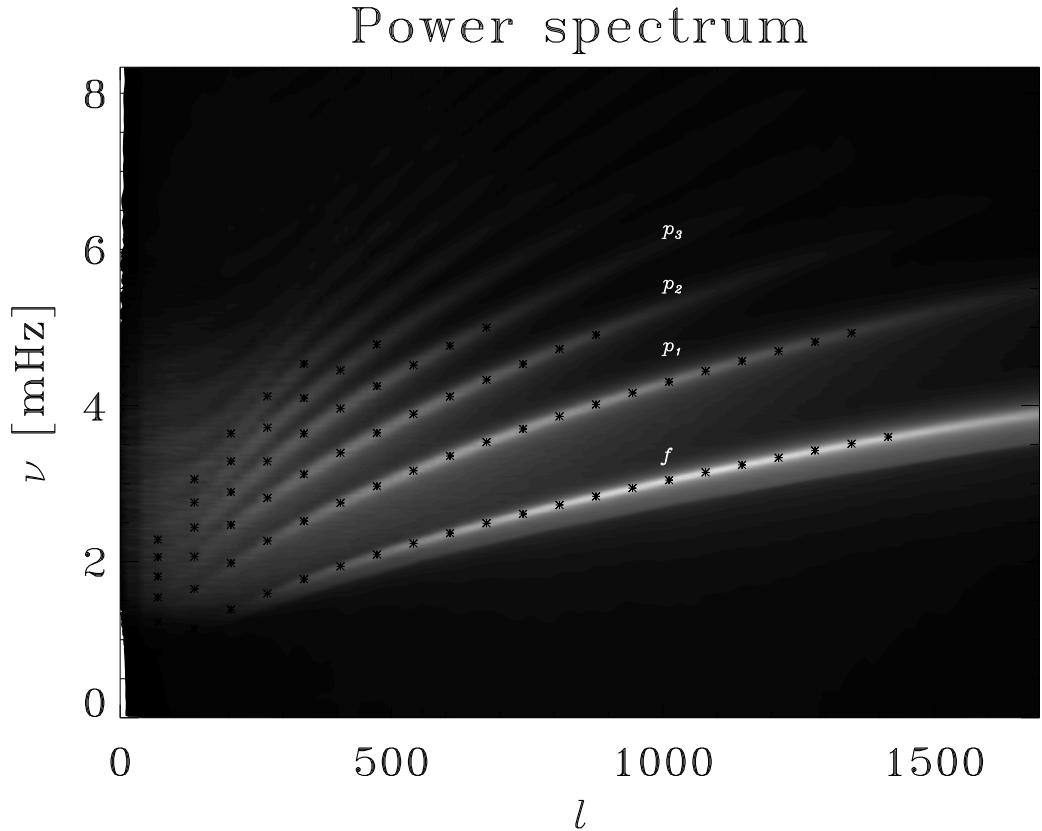}
\caption{Power spectrum obtained from a 12 hour `quiet' simulation. Some ridges have been labelled. The symbols mark independent estimates (obtained using MATLAB) of the eigenfrequencies of the altered solar model. The agreement between computation and theory appears reasonable.}
\label{power.spec}
\end{figure}

\begin{figure}[!ht]
\centering
\epsscale{1.0}
\plotone{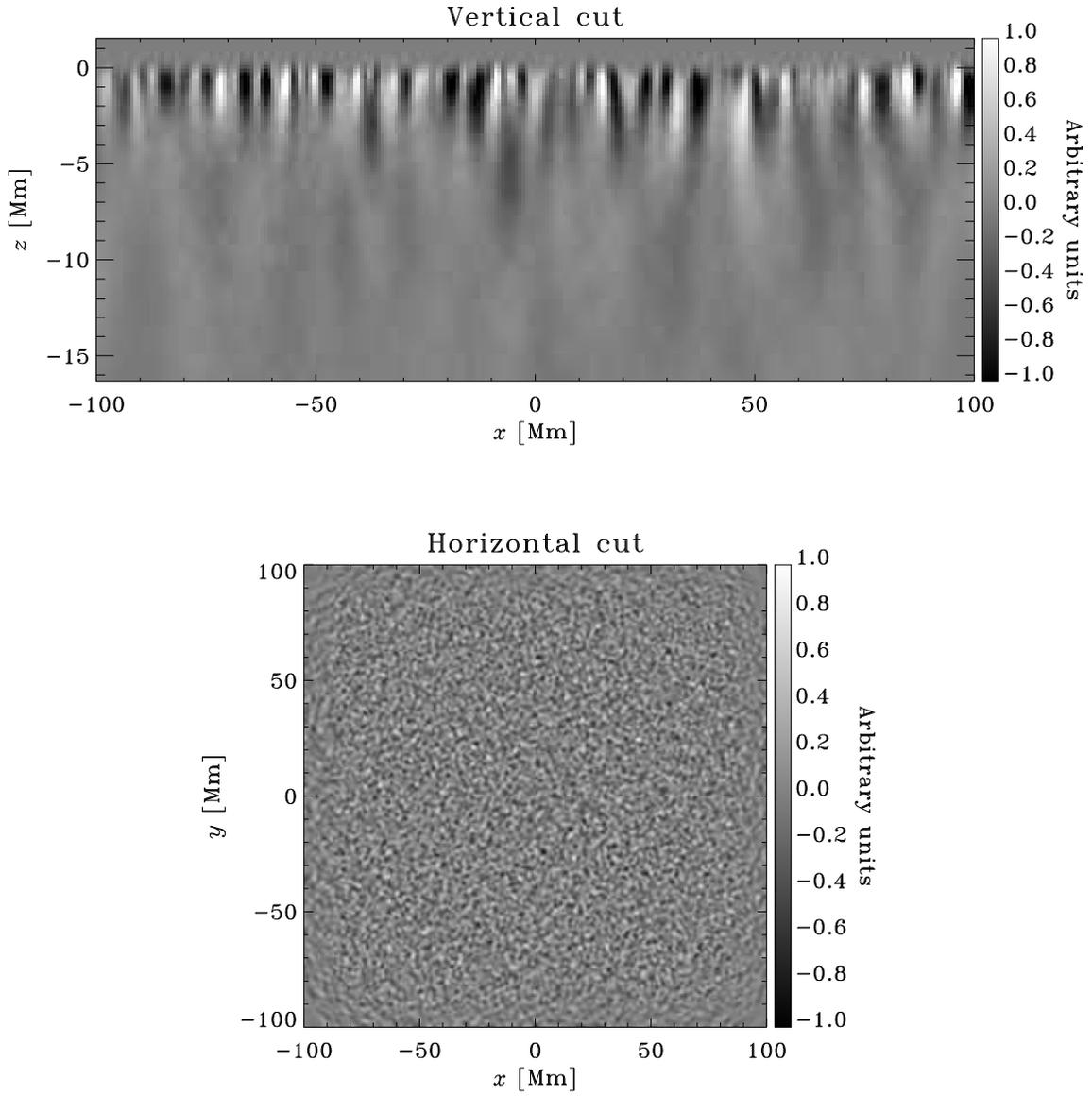}
\caption{Snapshots of the normalized vertical component of the oscillation velocity ($\sqrt{\rho_0 c} ~v_z$) - vertical and horizontal (at $z = 200$ km) cuts from a `quiet' simulation (absorbing boundary case) are displayed. The units are arbitrary and same for both panels. Energy conservation requires an increase in velocities to offset the sharply decreasing density in the near-surface layers - and hence the choice of this normalization ($\rho_0$ is the density and $c$ the sound speed).}
\label{vels.profile}
\end{figure}

The power spectrum and snapshots of the oscillation velocities derived from a 12 hour long `quiet' simulation are displayed in Figures~\ref{power.spec} and~\ref{vels.profile} respectively.  To simulate source suppression, the forcing term is multiplied by a spatial function that mutes source activity in a circular region of 10 Mm radius (i.e., the forcing function assumes a reduced value in this region). Based on estimates of emitted energy flux in sunspot umbrae, which range from 10 - 20\% of the average value in the quiet Sun \citep[e.g.,][]{schussler}, we perform two simulations, one with source strength in the disc region set to zero and another with 20\% of the `quiet' value. The two simulations possess very similar travel-time maps; therefore, we only show results from the simulation where the sources were completely suppressed.

The power spectral distribution of oscillation modes and the steady-state wave flux emerge from an interplay between source activity, wave damping, and mode mass. The non-scattering nature of source-strength perturbations complicates matters because the phase measurements are sensitive to the degree of inhomogeneity, which in turn is dependent on the intensity of the ambient wave flux. One can imagine that in the limit of an arbitrarily large wave flux, the time-shift effects of the suppressed source may altogether vanish (or reach some asymptotically small value). It is therefore important to investigate this matter in some detail. The wave flux in the computations is set by the choice of boundary conditions and damping rates. All other parameters being identical, absorbing horizontal sides evidently engender a weaker ambient flux than their periodic counterparts; thus we may study the impact of boundary conditions on the time shifts through numerical experiments with these choices of horizontal boundaries. In an indirect manner, these boundary conditions mimic higher and lower damping rates. In the case of the former, the lack of incoming waves from regions external to the boundaries sets the damping length (or maximum propagation distance) to approximately half the size of the computational region, since the perturbation is always at the center. This results in a dearth of $p$ modes that travel large distances or those that possess long lifetimes. Contrarily, in the case of periodic boundaries, waves exit from one boundary only to re-enter the domain from another; if the modal damping rates are unrealistically small, these perpetually propagating acoustic waves will rapidly fill up the domain, thereby significantly diluting the effects arising from the suppressed sources. Roughly we may conclude that the low wave damping limit is given by the periodic case and high damping limit by the absorbing case. 

Unfortunately, due to poor observational constraints on damping rates, it is unclear as to which of these situations is preferable. The linewidths recovered from Michelson Doppler Imager \citep[MDI][]{scherrer} observations by \citet{korzennik} and \citet{duvall98} differ by almost a factor of 2. Moreover, the complicated functional dependence of damping on frequency \citep[e.g.][]{korzennik} makes it all but impossible to implement it in the computation. Thus we can only hope to place bounds on the extent of the effect of suppressed sources since precise estimates are closely tied to the non-trivial feat of accurately matching the simulated wave power spectral distribution with the observational one. The outcomes of these tests are discussed in $\S$\ref{travel.times}.


\subsection{Theoretical model}\label{theoretical.model}
In order to gain an appreciation for the effects of damping on the conclusions of this paper, we create semi-analytical forward models in the manner of Gizon \& Birch (2002). These forward models predict the time shift associated with a specific perturbation. The starting point is the temporal Fourier transform of equation~(22) from Gizon \& Birch (2002), which gives the expected value of the cross-covariance, $C$, in terms of Green's functions ${\cal G}$ and the source covariance $M$,
\begin{equation}
C(\br_1,\br_2,\omega) = (2\pi)^2 \int\!\!\!\int d {\bf s} \; {\cal G}^{i*}(\br_1,\bs,\omega) {\cal G}^{j}(\br_2,\bs,\omega) M^{ij}(\bs,\omega) \, .
\label{eq.xcorr}
\end{equation}
The integration variable $\bs$ runs over the horizontal position of all the wave sources, $\br_1$ and $\br_2$ are the positions of the two observation positions, and $\omega$ is the temporal frequency.  Notice that in writing this equation we have assumed that the wave sources are uncorrelated in space.  In order to compute equation~(\ref{eq.xcorr}) we use the normal-mode summation of Green's functions from \citet{birch}, which include two models of wave damping, one based on the line-widths measured by \citet{korzennik} and the other with twice these rates \citep[approximately those measured by][]{duvall98}. We use the source covariance from \citet{birch}, though modified to include the reduction of source strength inside the disc of radius 10~Mm. It is important to note that the type of source used in this particular forward model is quadrupolar in nature, whereas we employ dipoles in the simulation. With these ingredients, the expected value of the point-to-point cross correlation (Eq.~[\ref{eq.xcorr}]) can then be computed numerically and averaged to obtain center-to-annulus cross correlations. In $\S$\ref{travel.times}, we shall further discuss the connection between the horizontal boundary conditions implemented in the simulations and the damping rates incorporated in this theory.




\section{TRAVEL TIMES AND POWER CORRECTION \label{travel.times}}
\begin{figure}[!ht]
\centering
\epsscale{1.0}
\plotone{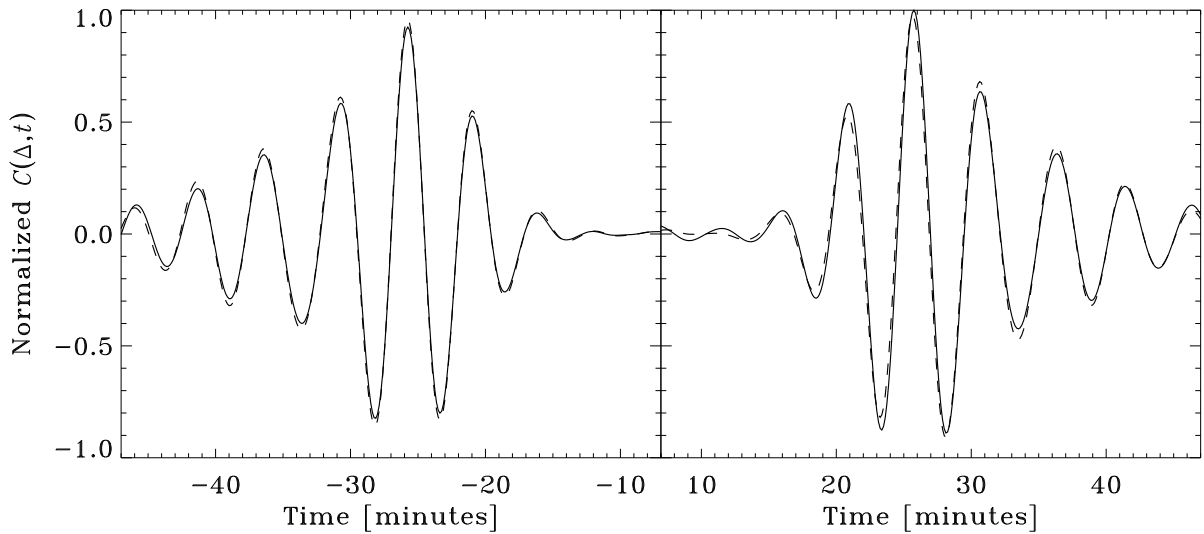}
\caption{Average cross correlation, $C(\Delta, t)$ for ingoing (on the left) and outgoing (on the right) waves obtained from a center-to-annulus scheme \citep{Duvall1996} for $\Delta = 24.35$ Mm from simulations with absorbing horizontal sides. The solid line shows the average cross correlation for a simulation with a spatially homogeneous source distribution (`quiet') and the dashed line for the source-suppressed region. The averaging is performed over a region within the 10 Mm disc in the quiet and perturbed simulations. The slight difference in amplitudes (there are no phase differences) between the in- and outgoing wave cross correlations in the quiet simulation is due to the absence of incoming waves from outside the boundaries. For the source-suppressed case, the outgoing wave cross correlation shows a phase advance (roughly 6 seconds) while the corresponding ingoing wave correlation shows a much smaller phase shift. This may contribute to the asymmetry between ingoing and outgoing waves observed in sunspots \citep[e.g.,][]{lindsey2005}.}
\label{cross_corr}
\end{figure}

\begin{figure}[!ht]
\centering
\epsscale{0.7}
\plotone{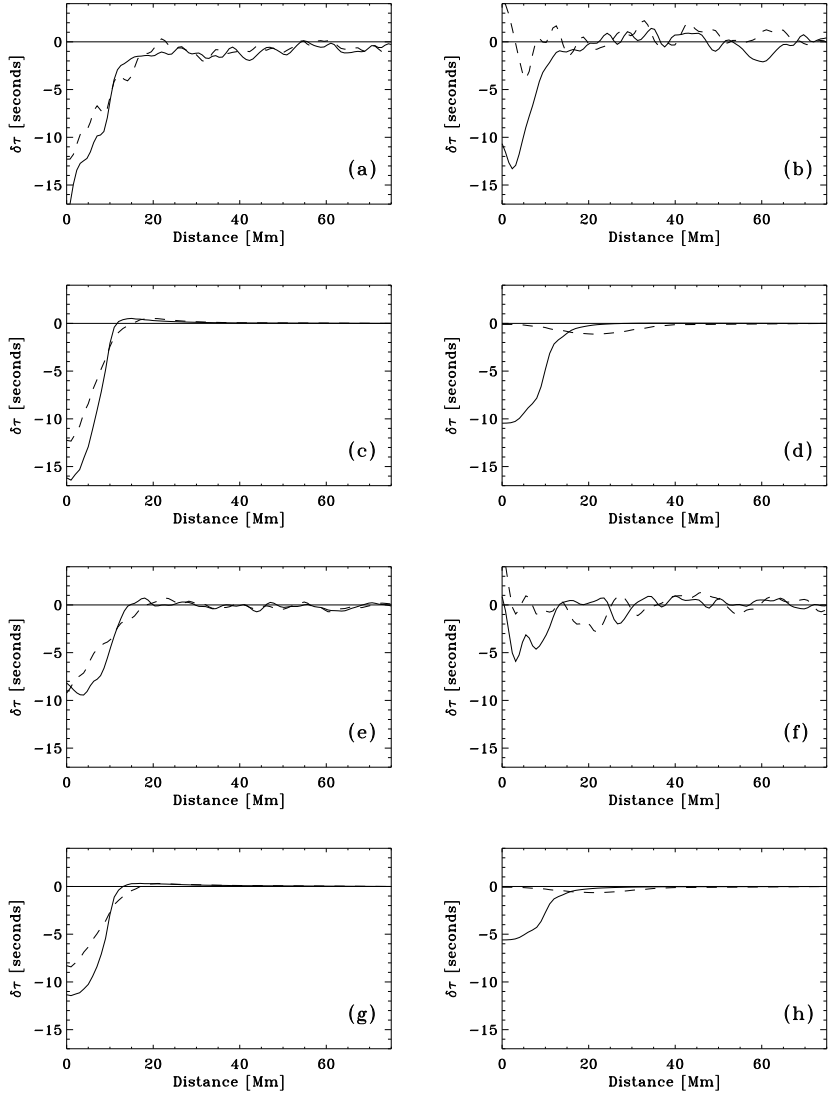}
\caption{Azimuthally averaged outgoing (solid line) and ingoing (dashed line) time shifts, $\delta\tau$, of waves that travel distances $\Delta = 6.2$ Mm (left column) and $\Delta = 24.35$ Mm (right column). The zero point is the center of the source suppression region. The first row (panels a, b) shows time shifts from the simulation with absorbing boundaries while the theory of $\S$\ref{theoretical.model} with high damping rates predicts those of the second row (panels c, d). The third row (panels e, f) is from the simulation with periodic boundaries while the bottom row (panels g, h) is from the theory with the \citet{korzennik} damping rates \citep[roughly half the linewidths of][]{duvall98}. A close correspondence is seen between the upper and lower pairs of rows.
\label{pmode_distances}}
\end{figure}

The $p$-mode travel times are measured using the procedure described in \citet{couvidat}. In order to estimate the travel times accurately, broad phase-speed filters were implemented to avoid contaminating the first bounce ridge with the filter artifact (see Table 1 of \citet{hanasoge07}; the FWHM was multiplied by 4). The $p$-mode cross-correlation branches corresponding to positive and negative times (outgoing/ingoing waves) averaged over the source-perturbation area in comparison to the average cross correlation for the quiet simulation with absorbing horizontal sides are shown in Figure~\ref{cross_corr} for $\Delta = 24.35$ Mm, where $\Delta = |\br_1 - \br_2|$ is the distance between measurement points. There is a noticeable asymmetry between the outgoing and ingoing waves, especially at larger distances where the outgoing waves contain almost all of the travel-time shift. Choosing the center of the source suppression to be the zero point, ingoing and outgoing travel-time shifts are azimuthally averaged and plotted in Figure~\ref{pmode_distances}. The reduction in the $p$-mode mean travel times seen in panel a of Figure~\ref{pmode_distances} is comparable, magnitude-wise, to the 15 s increase (azimuthal average) seen for $\Delta= 6.2$ Mm in some sunspots \citep[NOAA 8243, from high-resolution MDI observations,][]{couvidat}. The asymmetry between in and outgoing waves for the travel distance of $\Delta = 24.35$ Mm results in significant shifts in the difference times, of the order of 12 s (panel b, Figure~\ref{pmode_distances}).

In contrast, the simulations with periodic boundary conditions show reduced shifts \citep[also see][]{parchevsky07}, of the order of 10 s in the mean times for $\Delta = 6.2$ Mm and 6 s in the difference times for $\Delta =24.35$ Mm. Evidently, the reappearance of waves from the opposite boundary upon their exit from one has led to the prevalence of a larger wave flux in the computational domain. As pointed out in $\S$\ref{forward.model}, the wave flux plays a crucial role in bounding the effect of non-scattering source perturbations. The situation is rendered interesting by the close correspondence between the theory of $\S$\ref{theoretical.model} and the simulations, as seen between the upper and lower pairs of rows in Figure~\ref{pmode_distances}. Higher damping rates lead to larger time shifts and vice versa, analogous to simulations with the absorbing and periodic boundaries respectively. The conflicting linewidth estimates of \citet{korzennik} and \citet{duvall98} make it difficult to conclusively pick one simulation over the other. In fact, it is probably fair to say that realistic magnitudes of the time shifts lie somewhere between the estimates derived from the absorbing and periodic cases.

\begin{figure}[!ht]
\centering
\epsscale{1.0}
\plotone{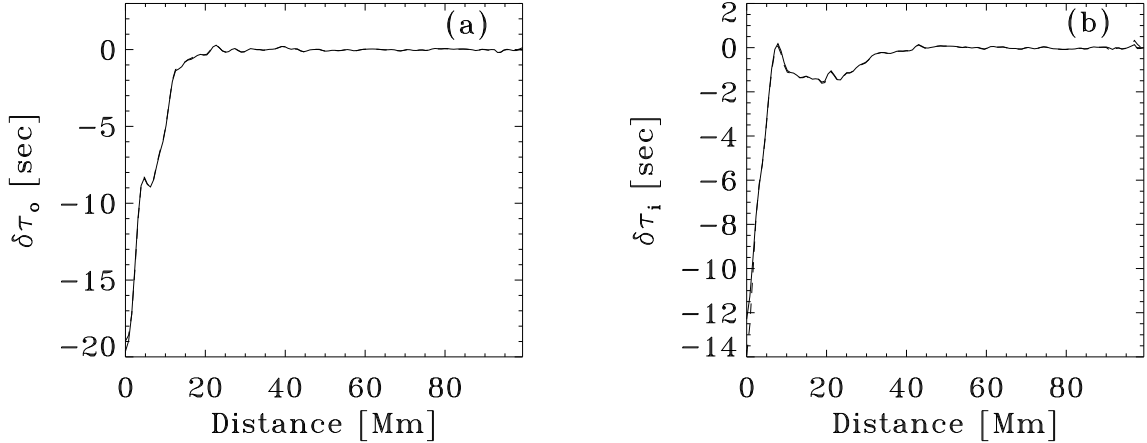}
\caption{Noise-subtracted \citep{hanasoge07} outgoing (panel a) and ingoing (panel b) time shifts for $\Delta = 24.35$ Mm. We performed three simulations, (I) sound-speed perturbation (of amplitude 7.5\% in $\delta c^2/c^2$ and size, 20 Mm) + source suppression, (II) only the sound-speed perturbation, and (III) only sources suppressed. The perturbations in II and III were identical to the individual components of I. The noise-subtracted travel times associated with I (solid line) is seen to be almost indistinguishable from II + III (dashed line), indicating that these wave field perturbations are entirely decoupled.\label{linearity}}
\end{figure}

If the simulations are believed to be representative of reality, and the travel times of in- and outgoing waves are appropriately `corrected' to account for the possibility of missing wave sources in sunspots, we might expect a significant change in the mean travel times for $\Delta = 6.2$ Mm. Moreover, the asymmetry between the in/outgoing waves seen for $\Delta=24.35$ Mm (ingoing $\sim$ -10 s, outgoing $\sim$ -40 s, azimuthal averages for sunspot NOAA 8243) could be reduced somewhat by applying these corrections. We show in Figure~\ref{linearity} that travel-time shifts associated with source suppression and sound-speed perturbations are linearly additive. Thus these source suppression effects can be addressed in a linear manner, making it possible to remove their signature from helioseismic analyses.


The decrease of acoustic power in a sunspot has been widely observed and explanations offered \citep[e.g.][]{lites}; recently, \citet{parchevsky} have suggested that the suppression of convection (and hence wave sources) is sufficient to explain more than half of the decrease in acoustic power in sunspots. However, in our calculations, even when the source strengths in the region of suppression are reduced to zero, we see only about 20\% reduction in acoustic power. In any case, it is difficult to compare these two results because of the differences in damping rates, the time length of the calculations, the sizes of the computational domains, etc. To compensate for travel-time measurement `errors' related to the local reduction in acoustic power, \citet{rajaguru} proposed a power correction method which we incorporated before computing travel times. Since we use broad phase-speed filters and because the acoustic power is reduced by only 20\%, the power correction does little in the way of altering time shifts ($\sim$ 10 \% change at the most) in our simulations. 

\begin{figure}[!ht]
\centering
\epsscale{1.0}
\plotone{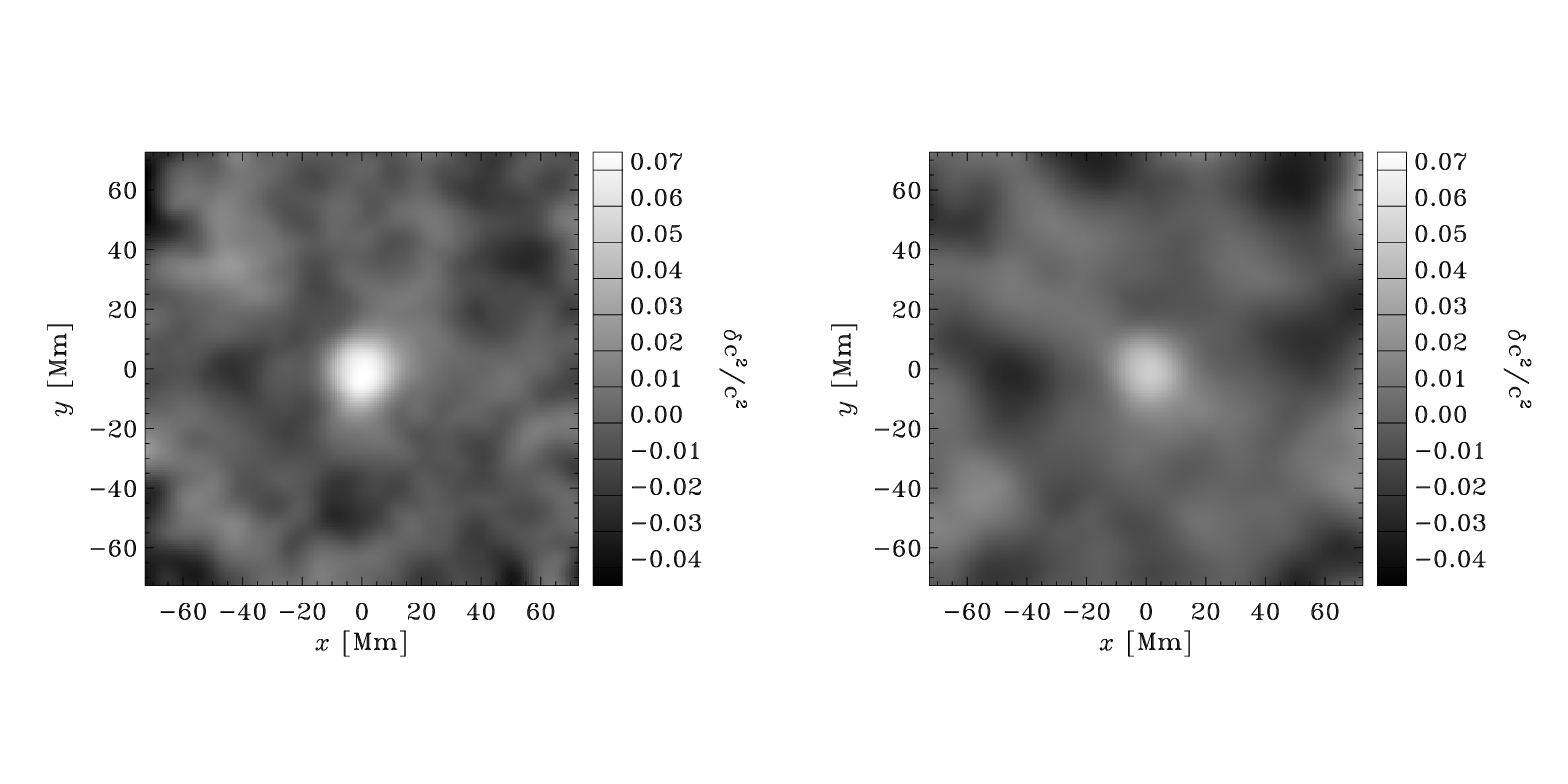}
\caption{Inversion of the noise-subtracted \citep{hanasoge07} mean time shifts arising from the suppressed sources. Shown are piece-wise constant slabs, averaged over the depth range [-0.62, 0] Mm (left panel) and [-1.42, -0.62] Mm  (right panel), where 0 indicates the surface. Because the inversion is noisy, the appearance of random features is observed. Travel-time signatures of suppressed sources and local increases in the sound speed look unexpectedly identical, showing significant cross talk from one effect onto the other.\label{inversion}}
\end{figure}

Inversions of the mean time shifts (absorbing boundary case) using sound-speed kernels in the ray approximation and the multi-channel deconvolution algorithm \citep{jensen} are shown in Figure~\ref{inversion}. The perturbation appears as a shallow ($\approx$ 1.5 Mm, abutting the photosphere), 7.5 \% increase in $\delta c^2/c^2$, where $c$ is the sound speed. The horizontal size of the anomaly is comparable to that of the region of 
suppressed sources, i.e. about 20 Mm.



\section{DISCUSSION}\label{discussion}
We demonstrate that obtaining meaningful travel times is strongly incumbent upon the homogeneity of sources in the medium. Numerical and analytical experiments where sources sources were suppressed over a region typically the horizontal size of a sunspot predict significant wave phase shifts. Therefore, our analysis seems to indicate that helioseismic investigations into the internal constitution of a sunspot are incomplete without taking into account the effects of inhomogeneously distributed sources and damping \citep{woodard}. We see that in- and outgoing waves are differentially affected, with the asymmetry exacerbated at increasing travel distance, $\Delta$, especially when damping rates are high. The large negative mean travel-time shifts seen at the shortest travel distances ($\sim$ -10 $-$ -15 s, $\Delta = 6.2$ Mm) are worrisome for it is not clear how accurate estimates are of the amplitude of the near-surface sound-speed perturbation below a sunspot. Similarly, the systematic difference travel times observed for waves (also $\sim$ -6 $-$ -15 s, $\Delta = 24.35$ Mm) that travel larger distances indicates that the flow inversions may be inaccurate. The power correction of \citet{rajaguru} in this case decreases the magnitude of the travel-time shifts at most by about $10 \%$. The sensitivities of other methods of helioseismology like ring diagram analysis \citep{hill} and acoustic holography \citep{lindsey97} to the homogeneity of the wave field remain to be investigated.

The upshot of these calculations is that the methods of \citet{Gizon2002} and \citet{birch} can be applied to infer and model out to a large extent the measurement biases introduced by the suppressed sources. Moreover, numerical forward models of the solar wave field have become increasingly sophisticated \citep[e.g.,][]{parchevsky, cameron07, hanasoge_mag}, presenting ways to test inversion results. 

\appendix

\section{Altered solar model}
\label{alt.model}
 Here, we describe the artificially convectively stabilized model \citep[Appendix A of][]{hanasoge1} used in our computations. The dimensionless radial co-ordinate is denoted by $r$, where $r$ expresses fractions of the solar radius $R_\odot = 6.959894677 \times 10^{10}$ cm. For $r < 0.98$, background properties as prescribed by model S \citep{jcd} are used. In the range $0.9998 \ge r \ge 0.98$, the empirical formulae:
\begin{eqnarray}
\rho_0 &=& 4.1522194 \left[0.998989 - r + 4.36138(r-0.98)^{2.1}\right]^{2.009828}, \\
p_0 &=& 2.7392767 \times 10^{15} \left[0.998989 - r + 4.36138(r-0.98)^{2.1}\right]^{3.009828}, \\
g &=& -\frac{1}{\rho_0 R_\odot} \frac{dp_0}{dr},\\
\Gamma_1 &=& \max(\Gamma^S_1,1.507550),
\end{eqnarray}
are implemented, whereas in the region $1.002 \ge r \ge 0.9998$, an isothermal layer is utilized:
\begin{eqnarray}
\rho_0 &=& 4.5260638 \times 10^{-7} \exp[7690.7995 (0.9998 - r)]\\
p_0 &=&  1.0252267 \times 10^{5} \exp[7690.7995 (0.9998 - r)] \\
g &=& 24998.23
\end{eqnarray}
Density ($\rho_0$) is expressed in units of ${\rm g~ cm^{-3}}$, pressure ($p_0$) in ${\rm dynes~ cm^{-2}}$, gravity ($g$) in ${\rm cm~ s^{-2}}$, the first adiabatic index ($\Gamma_1$) is dimensionless,  and the sound speed ($c$) in units of ${\rm cm~ s^{-1}}$ is given by:
\begin{equation}
c = \sqrt\frac{\Gamma_1 p_0}{\rho_0}.
\end{equation}
The eigenfrequencies of the altered model have been computed independently using a boundary value solver provided in MATLAB and compared with those recovered from the computations (Figure~\ref{power.spec}). The agreement is good.

\section{Code verification}
\label{verify}
The accuracy of the numerical scheme described in $\S$\ref{forward.model} is confirmed using a number of tests \citep{hanasoge_thesis}. Before delving into the verification details, it is important to understand the parameter regimes of the waves and the limiting factors controlling the simulation timestep. The highest frequency of waves of interest to us is of the order of 6 mHz, corresponding to a timescale of about 167 seconds. The simulation timestep of 2 seconds is significantly smaller than the period of the oscillations. The calculations are evidently temporally highly over resolved; compared to the 4-10 points per wavelength (ppw) quoted by \citet{hu} and \citet{berland}, the simulations operate at between 80-250 ppw. In the radial direction, the eigenfunctions of the modes contain a rather small number of nodes (10 - 30 depending on the mode) in comparison to the actual number of grid points (300 points). The reason for the excessive spatial resolution is the need to capture the rapid density (pressure) variation with radius. Therefore, the limiting factor controlling the timestep is the large number of density (pressure) scale heights in the computational domain, which is why the radial and temporal resolutions are so high.
 
\begin{figure}
\begin{center}
\plotone{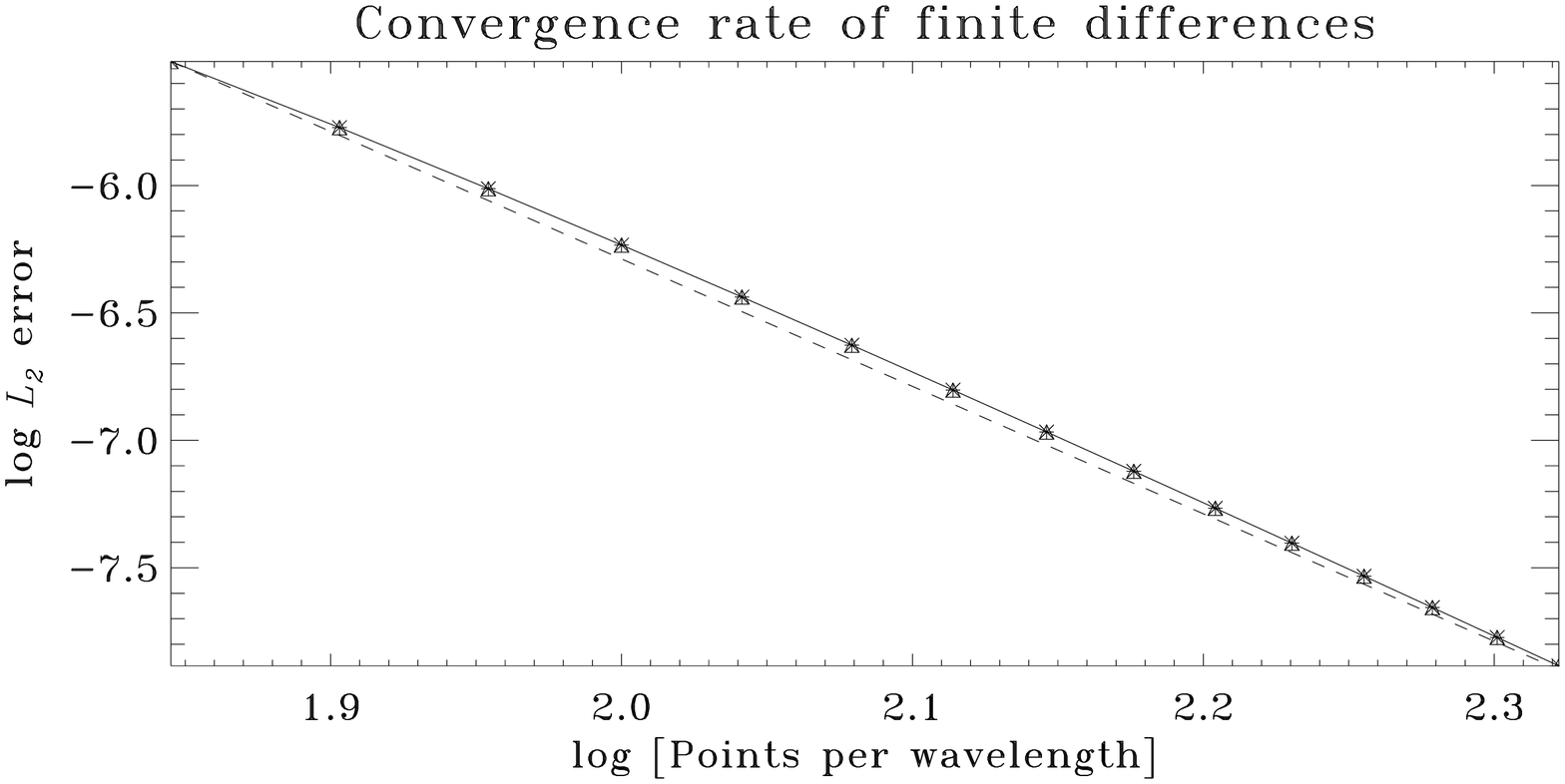}
\end{center}
\caption{Spatial convergence rate of the compact finite differences with fifth-order accurate boundary conditions. The solid line shows the accuracy of the scheme, while the dashed line is the theoretical fifth-order accuracy curve.}
\label{crude.conv}
\end{figure}

\begin{figure}
\begin{center}
\includegraphics[scale=0.45]{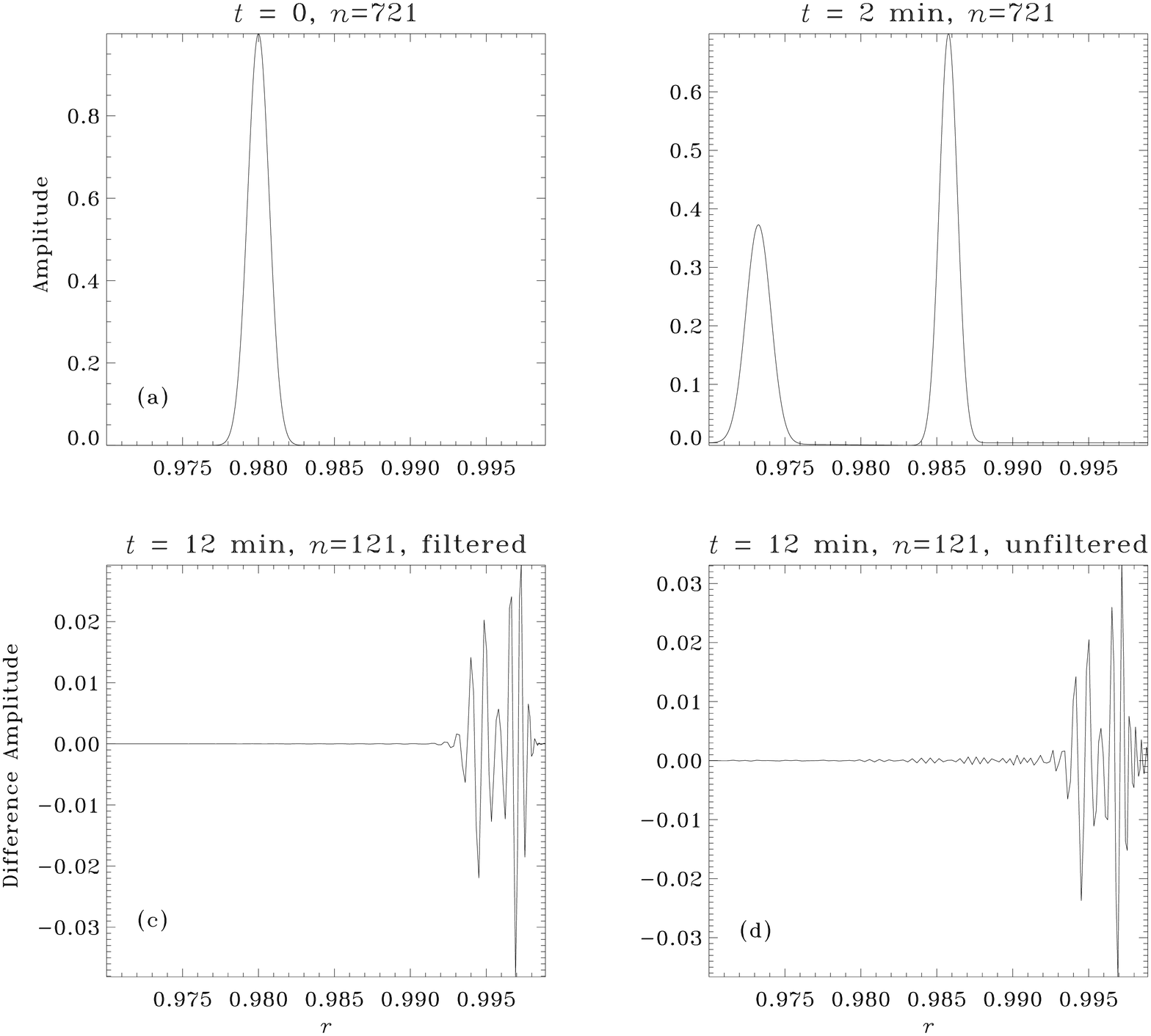}
\end{center}
\caption{Experiment to determine the spatial error convergence rate. The initial condition, a Gaussian wavelet in velocity, is shown in panel (a). In (b), the temporally evolved wavelet at time $t = 2$ min is displayed. Simulations are performed with varying numbers of grid points, $n=721, 361, 181, 145,$ and 121, so that each grid is a downsampled version (i.e., every other point, every third point etc.) of the $n=721$ case. Errors are computed at $t = 2$ min using a downsampled version of the $n=721, t = 2$ min solution as a template (panel b). In panels (c) and (d), the differences between the $n=121$ solution and the downsampled $n=721$ template at $t=12$ min are displayed. The wavelet has propagated all the way out to the upper boundary at this point; it is seen that the difference, interpreted as the error, is greater in the unfiltered case in panel (d) than in the filtered version in panel (c), where the filter is applied to dealias variables in the radial direction \citep{dealias}. The difference between (c) and (d), which although appears harmless, continues to grow, eventually overwhelming the simulation unless a de-aliasing filter is applied frequently. }
\label{experiment}
\end{figure}

\begin{figure}
\begin{center}
\includegraphics[scale=0.4]{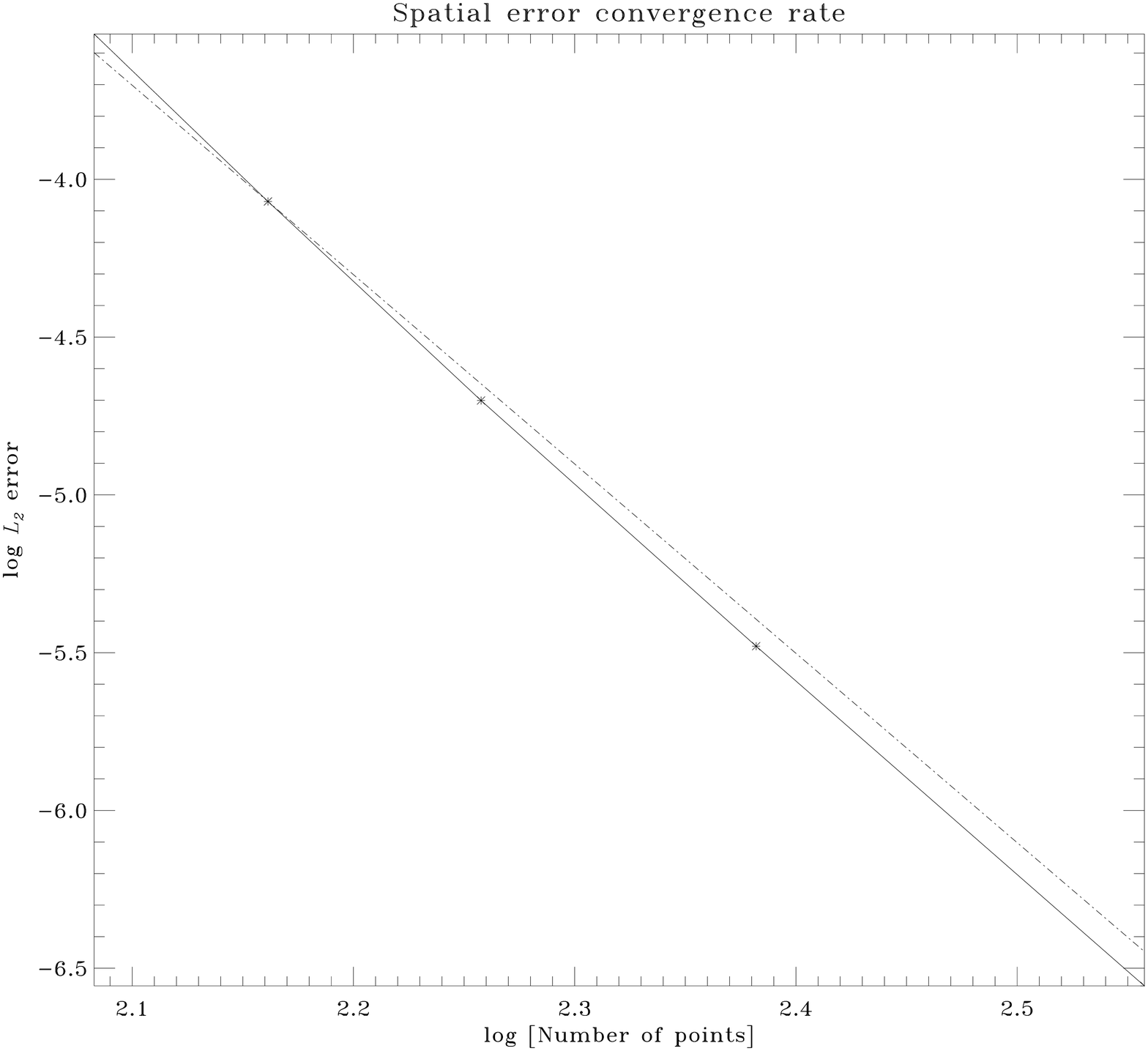}
\end{center}
\caption{Spatial error convergence rate (with radial dealiasing) based on the experiment of Figure~\ref{experiment}; the time step was $\Delta t = 0.05$ seconds. The solid line is the error of the compact finite differences and the dashed line is a theoretical sixth-order accuracy curve. It is somewhat surprising that the scheme obeys a sixth-order accuracy law despite the use of fifth-order boundary conditions. Partly, the reason could be that the problem is a consistent initial-boundary value problem, i.e. $v =0$ and $\partial_z p = -\rho g$ at the boundaries.}
\label{space.conv}
\end{figure}

We show in Figure~\ref{crude.conv} that the boundary conditions cause the error convergence rate of the compact finite differences to drop to fifth order. Although not presented here, the convergence rate is entirely unchanged when the radial de-aliasing filter, described in \citet{dealias}, is applied in conjunction with the finite differences. Next, to demonstrate the accuracy of the spatial scheme in its entirety (i.e., when used with radial de-aliasing and the temporal scheme), we simulate the 1-D propagation of a Gaussian wavelet in a box with reflecting boundary conditions. The grid-spacing in the calculation follows the constant travel-time criterion developed in \citet{hanasoge1}. The background model is chosen to be an adiabatically stratified, truncated polytrope with index $m=1.5$, gravity ${\bf g} = - 2.775 \times 10^4~{\rm cm~s^{-2}}{\bf e_z}$, reference pressure $p_{ref} = 1.21 \times 10^5~{\rm dynes~cm^{-2}}$ and reference density $\rho_{ref} = 2.78 \times 10^{-7}~{\rm g~cm^{-3}}$, such that the pressure and density variations are given by,

\begin{equation}
p_0(z) = p_{ref}\left(-\frac{z}{z_0}\right)^{m+1},
\label{backk.pressure}
\end{equation}
and
\begin{equation}
\rho_0(z) = \rho_{ref}\left(-\frac{z}{z_0}\right)^{m}.
\label{backk.density}
\end{equation}
The photospheric level of the background model is at $z=0$, with the upper boundary of the truncated polytrope placed at a depth of $z_0 = 768~{\rm km}$. This model is similar to the stratification in the outer layers of the Sun \citep[e.g.,][]{bogdan95}. Because error convergence rates are very sensitive and easily masked by slight differences such as the locations of the comparison points of solutions, we start with a highly resolved 721 point grid and downsample by successively higher rates (every second point, every third point, and so on). The solutions obtained on this sequence of grids are compared with the highly resolved case to obtain the error convergence rate. The lower boundary of the simulation is placed at $z = -20.876$ Mm, with wall-like boundary conditions on both ends ($v=0, \partial_z p = -\rho g$, at the boundaries). The timestep of the simulation was chosen to be $\Delta t = 0.05$ seconds. The experiment is graphically displayed in Figure~\ref{experiment} and the error convergence rate is shown in Figure~\ref{space.conv}.

\subsection{Eigenfunctions}\label{eig.funcs}
\begin{figure}
\begin{center}
\includegraphics[scale=0.75]{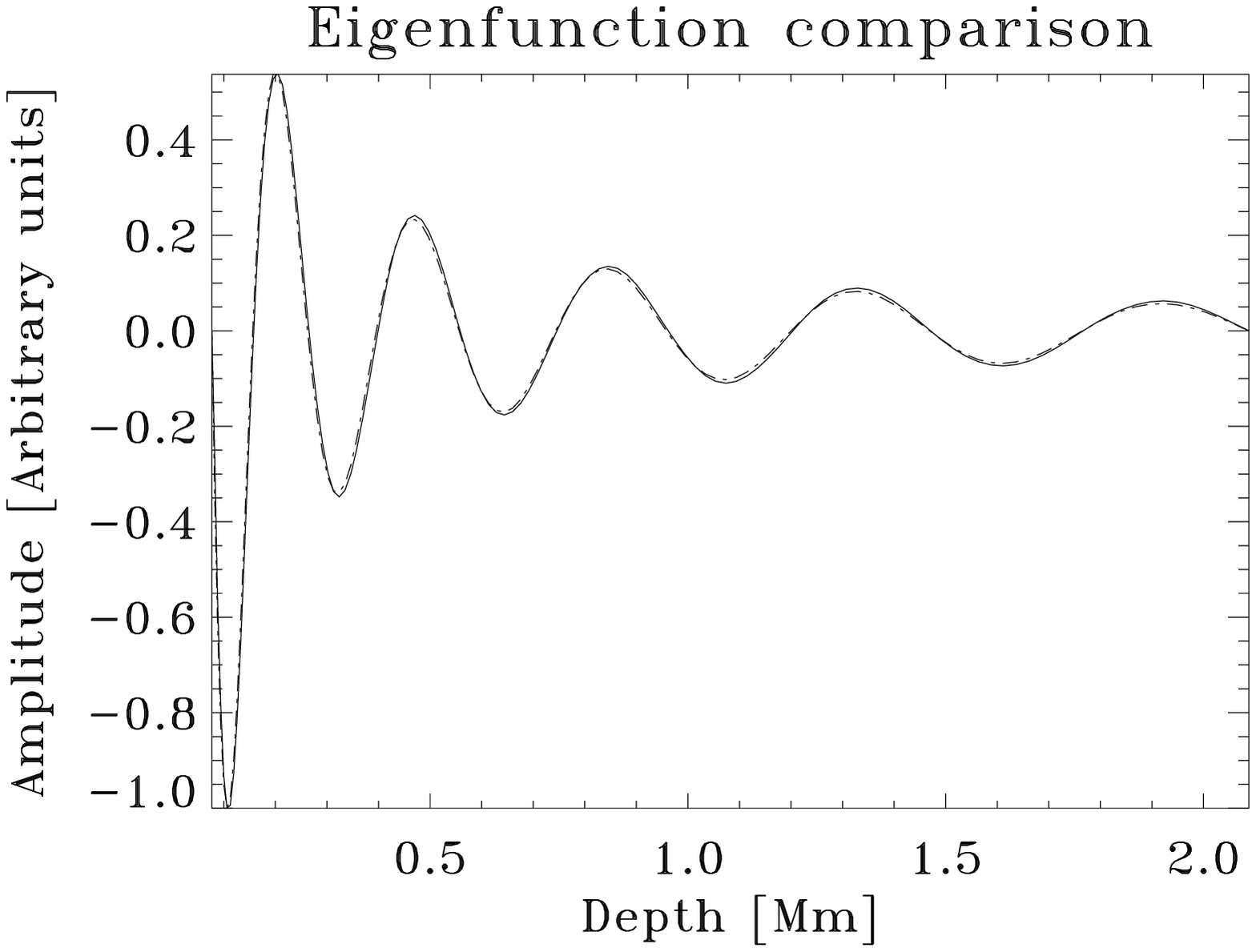}
\end{center}
\caption{Comparison of eigenfunctions for a resonant mode of frequency $\nu = 6.6111$ mHz, obtained analytically (solid line) and through simulation (dot-dash line) with $n = 121$. At higher resolutions, the two curves are virtually indistinguishable and hence are not shown here. Eigenfunctions for other resonant frequencies have also been compared and found to be in good agreement. Including the two boundaries, the eigenfunction contains only eleven nodes, far smaller than the number of grid points. With fewer ($\lesssim 80$) points, the system develops instabilities because of the steep density gradient.}
\label{eigfunc.comp}
\end{figure}

For the polytrope described above, it is possible to determine the eigenfunctions analytically \citep[e.g.,][]{bogdan95}. This will assist us in verifying that the spatial scheme is able to recover the eigenfunctions accurately. The first step is to set down the equations to be solved:
\begin{eqnarray}
\partial_t\rho(z,t) &=& -\partial_z (\rho_0 v)\label{contt}\\
\rho_0 \partial_t v(z,t) &=& -\partial_z p -\rho g\label{momm}\\
\partial_t p(z,t) &=& -c^2_0 \rho_0 \partial_z v + \rho_0 v g\label{energyy},
\end{eqnarray}
where $\rho$ refers to density, $c$ refers to sound speed, the 0 subscript refers to background properties of the model, and $t$ time. Differentiating equation~(\ref{momm}) with respect to time and substituting for time derivatives of density and pressure from equations~(\ref{contt}) and~(\ref{energyy}) respectively, we obtain the following:
\begin{equation}
\rho_0 \partial^2_t v(z,t) = -\partial_z (-c^2_0 \rho_0 \partial_z v + \rho_0 v g)+\partial_z (\rho_0 gv).\label{simplified.1}
\end{equation}
Next we define the Eulerian pressure and velocity fluctuations to be, respectively:
\begin{eqnarray}
p(z,t) &=& \pst(z) e^{-i \omega t}\label{predef.1}\\
v(z,t) &=& \vst(z) e^{-i \omega t}\label{veldef.1}.
\end{eqnarray}
Substituting these expressions into equation~(\ref{simplified.1}), we have:
\begin{equation}
- \omega^2 \rho_0 z_0^2 v^*  = \partial_s (c^2_0 \rho_0 \partial_s v^*), \label{midway.1}
\end{equation}
where once again, $s= -z/z_0$, $\rho_0 = \rho_c s^m$, $p_0 = p_c s^{m+1}$, $c^2_0 = \tilde{c^2} s$, and $\rho_c,p_c,\tilde{c^2}$ are the density, pressure and sound speed square at $s=1$. The upper and lower boundaries of the polytrope are at spatial locations $s=1, D$, with $D$ a free parameter. Equation~(\ref{midway.1}) is simplified to obtain:
\begin{equation}
s\partial^2_s\vst + (m+1)\partial_s\vst + \frac{\alpha^2}{4}\vst=0, \label{solve.this.1}
\end{equation}
where $\alpha = 2\omega z_0/\tilde{c}$. Equation~(\ref{solve.this.1}) is solved to obtain the analytical expression for the eigenfunction: 
\begin{equation}
\vst = A s^{-m/2} J_m(\alpha s^{1/2}) + B s^{-m/2} Y_m(\alpha s^{1/2}).\label{eigens}
\end{equation}
The constants $A$ and $B$ are determined by enforcing the boundary conditions $\vst(s=1) =0$ and $\vst(s=D) = 0$. From these conditions emerge a sequence of resonant frequencies, $\alpha$, which can then be used to obtain the eigenfunctions of the resonant modes. The eigenfunction for pressure is related to the one for velocity according to:
\begin{equation}
\pst = \frac{2i\rho_c\tilde{c}}{\alpha}s^m[m\vst + s\partial_s\vst].\label{eigens.press}
\end{equation}
 To obtain eigenfunctions from the computations, we excite waves and simulate wave propagation in the above-described cavity. Temporal transforms of the entire dataset are computed at each spatial location; resonant modes are then isolated by analyzing large amplitude regions in the power spectrum. These frequencies are compared to the analytically predicted values to ensure that these are indeed resonant modes. Having done so, the temporal spectrum is multiplied by a frequency-window function to retain power only in the region of interest and then inverse Fourier transformed. The resultant transforms are the desired eigenfunctions. However, spatial error convergence rates are difficult to measure from this experiment because the eigenfunction signal is diluted by neighboring modes due to the finite temporal window of the simulations. Moreover the accuracy with which the resonant frequency can be measured is bounded by the time length of the calculation. For the eigenfunction shown in Figure~\ref{eigfunc.comp}, a resonant mode with $\nu = 6.6111$ mHz was isolated using an extremely narrow, four-point box-car frequency filter. Simulations with varying grid spacings all showed a peak in the power spectrum at frequency of 9 $\mu$Hz away from the analytical prediction (frequency resolution $\sim 22 \mu$Hz, from a 12-hour simulation).

\subsection{Efficacy of the transmitting boundary}
As described in \citet{hanasoge1}, we use the transmitting boundary conditions of \citet{thompson} with an adjoining sponge \citep[e.g.,][]{lui} to `prepare' the waves for the boundary. The main reason for using this prescription as opposed to other possibilities (\citet{poinsot}; see \citet{colonius} for a review) is the ease of implementation and efficiency of the method. In the simulations, we use the following:
\begin{eqnarray}
\frac{\partial p}{\partial z} |_{z = {\rm bot}} &=& -~ c \rho_0 \frac{\partial v_z}{\partial z} - \rho g,\label{bc1.eq} \\
\frac{\partial p}{\partial z} |_{z = {\rm top}} &=& +~ c \rho_0 \frac{\partial v_z}{\partial z} - \rho g, \\
\frac{\partial p}{\partial x} |_{x = {\rm left}} &=& -~ c \rho_0 \frac{\partial v_x}{\partial x},\\
\frac{\partial p}{\partial x} |_{x = {\rm right}} &=& +~ c \rho_0 \frac{\partial v_x}{\partial x}\label{bc4.eq},
\end{eqnarray}
with the velocity derivatives computed in a Dirichlet sense, using the values at the end points.

To test if these boundary conditions change the eigenfunction in any significant manner and to ensure that to a large extent, they are indeed non reflecting, we perform 1D calculations of wave propagation in a background similar to that of $\S$ \ref{eig.funcs}. 
To give this problem a realistic spin, we stitch an isothermal atmosphere to the polytrope so that a finite acoustic cut-off frequency is achieved, thereby providing a natural reflection region for the waves. Moreover, we relax the zero-velocity condition on the upper boundary and implement a combination of the sponge and transmitting boundary conditions (Eqs. [\ref{bc1.eq}] - [\ref{bc4.eq}]) while still enforcing a zero-velocity condition on the lower boundary. Waves whose frequencies are lower than the acoustic cutoff are reflected back into the interior while an evanescent non-propagating region develops in the isothermal atmosphere. Thus, we can determine the effect of the boundary conditions on the simulated eigenfunctions by comparing them with their analytical counterparts.

\subsection{Evanescent behavior}
Consider an isothermal layer with constant sound-speed $c_0$ with exponentially decaying background density ($\rho_e$)and pressure ($p_e$) profiles smoothly connected to the truncated polytrope of \S \ref{eig.funcs}. We have:
\begin{eqnarray}
\rho_e &=& \rho_{ref} e^{-(z_0+z)/H},\\
p_e &=& p_{ref} e^{-(z_0+z)/H},\\
T_e &=& T_{ref},
\end{eqnarray}
with $z=0$ corresponding to the `photosphere' of this model, and $H$ to the scale height in the atmosphere.
 The governing equation~(\ref{midway.1}) is unaltered except for the background density and sound speed. Again, we define $v(z,t), p(z,t)$ as:
\begin{eqnarray}
p(z,t) &=&  \pst_e(z) e^{-i\omega t},\\
v(z,t) &=& \vst_e(z) e^{-i\omega t}.
\end{eqnarray}
When the constituent equation~(\ref{midway.1}) is solved, we obtain the following for $\pst_e$ and $\vst_e$:
\begin{eqnarray}
\pst_e &=&  C e^{\lambda z -z/H},\\
\vst_e &=&  D e^{\lambda z},
\end{eqnarray}
with constants $C$, $D$ and $\lambda$ a solution of:
\begin{eqnarray}
\lambda^2 - \frac{\lambda}{H} + \frac{\omega^2}{c_0^2}&=& 0,\\
\lambda &=& \frac{1}{2H}\left[1 - \sqrt{1 -\frac{\omega^2}{\cut^2}}\right],\\
\cut &=& \frac{c_0}{2H}.
\end{eqnarray}
We retrieve two solutions for $\lambda$ and reject the one whose energy density $\propto \rho v^2$ grows without bound as a function $z$. In this situation, the relation between $p^*_e$ and $v^*_e$ is given by:
\begin{eqnarray}
v^*_e &=& \frac{i\omega}{\rho_c\eta}p^*_e, \label{another.relation}\\
\eta &=& c_0^2\lambda - g.
\end{eqnarray}

For boundary conditions, we use normal velocity and Eulerian pressure matching across the boundary $s=1$:
\begin{eqnarray}
\vst &=& v^*_e,\\
\pst &=& p^*_e,
\end{eqnarray}
where $\vst$ and $\pst$ are the velocity and pressure in the polytropic layer, given by equations~(\ref{eigens}) and~(\ref{eigens.press}) respectively. When writing the velocities in the following form, we will have only the pressure equation to solve:
\begin{eqnarray}
\vst &=& A\frac{i\omega}{\rho_c\eta} e^{-\lambda z_0} s^{-m/2}[J_m(\alpha s^{1/2}) + \beta Y_m(\alpha s^{1/2})]\label{eig.int},\\
v^*_e &=& A\frac{i\omega}{\rho_c\eta}e^{-\lambda sz_0}[J_m(\alpha) + \beta Y_m(\alpha)],\label{rand.eq}
\end{eqnarray}
where $\beta$ is the unknown constant we must determine ($A$ can be arbitrarily chosen). Equations~(\ref{another.relation}) and~(\ref{rand.eq}) constrain $p^*_e$:
\begin{equation}
p^*_e = Ae^{-\lambda sz_0 +sz_0/H}[J_m(\alpha) + \beta Y_m(\alpha)].
\end{equation}
 Matching $p^*_e = \pst$ at $s=1$ gives us the following relations:
\begin{eqnarray}
\beta &=& -\left[\frac{J_m(\alpha)+\kappa J_{m-1}(\alpha)}{Y_m(\alpha)+\kappa Y_{m-1}(\alpha)}\right],\label{eq.beta}\\
\kappa &=&\frac{\omega c_0}{\alpha\eta}e^{-z_0/H}.
\end{eqnarray}
To determine the resonant modes $\alpha$ of this model, we use the definition of $\beta$ from equation~(\ref{eq.beta}) and set equation~(\ref{eig.int}) to zero at $s= D$. Having then recovered the resonant frequencies, the pressure and velocity eigenfunctions in the interior ($s \le 1$) may be obtained by evaluating:
\begin{eqnarray}
\vst &=& A\frac{i\kappa}{\rho_c\tilde{c}} e^{-\lambda z_0 + z_0/H} s^{-m/2}[J_m(\alpha s^{1/2}) + \beta Y_m(\alpha s^{1/2})],\\
\pst &=& -A\kappa e^{-\lambda z_0 + z_0/H} s^{(m+1)/2}[J_{m-1}(\alpha s^{1/2}) + \beta Y_{m-1}(\alpha s^{1/2})].
\end{eqnarray}

The acoustic-cutoff frequency, $\omega_c$, of the model ($D\ge s\ge 1$) is given by:
\begin{equation}
\omega_c = \frac{c_0\sqrt{m^2+1}}{2z_0}\frac{1}{\sqrt{s}}.
\end{equation}

\begin{figure}
\begin{center}
\includegraphics[scale=0.75]{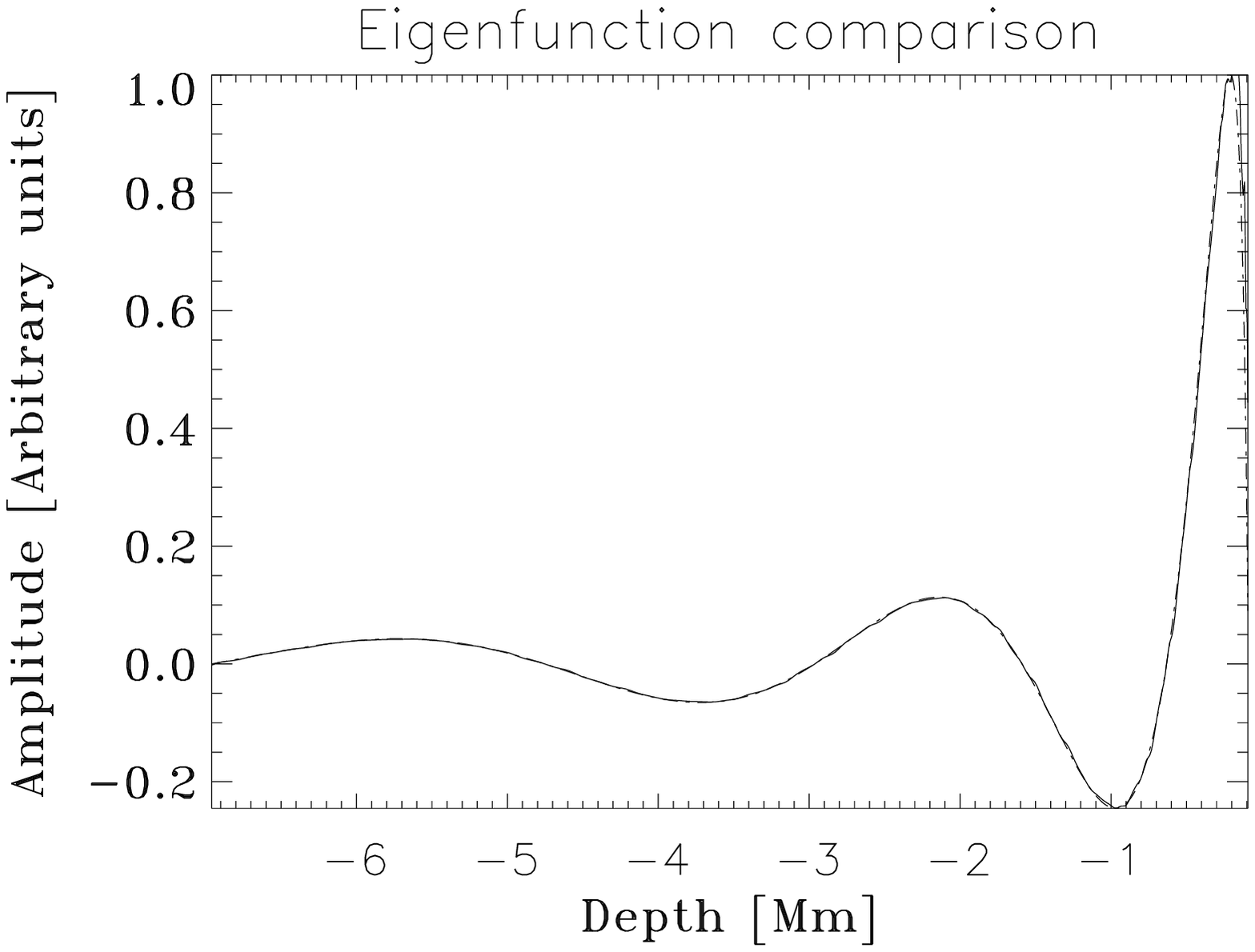}
\end{center}
\caption{Simulated (solid line) and analytical (dot-dash line) eigenfunctions for $\nu = 1.68$ mHz, for the model described above. It is seen that the boundary conditions and sponge do not affect the eigenfunction over the region of interest; although there is an amplitude error of a few \% in the upper-most layers of the polytrope, the interior nodes are oblivious to the boundary conditions. This eigenfunction was obtained from a 24-hour simulation wherein the waves were constantly excited over a small region in the interior.}
\label{test.eig.bc}
\end{figure}

The model for this particular test is parametrized by $m=1.5$, $z_0 = 768$ km, $D=90.6198, c_0 = 8.51$ km ${\rm s^{-1}}$, $p_0 = 1.21 \times 10^5$ dynes ${\rm cm^{-2}}$, $\rho_0 = 2.78 \times 10^{-7}$ g ${\rm cm^{-3}}$, $H = z_0/(m+1)$ km, and $g = 14160. \times 10^5$ cm ${\rm s^{-2}}$. Plotted in Figure~\ref{test.eig.bc} are the analytical (dotted line) and the simulated (solid line) eigenfunctions. The sponge is placed adjacent to the upper boundary (located 1232 km above $z_0$). As can be seen the presence of the sponge does not affect the interior parts of the acoustic eigenfunction. There is an amplitude error near the upper-most region of the polytrope due to the combined influence of the boundary condition and the sponge but the nodes remain mostly unaffected.

\begin{figure}
\begin{center}
\includegraphics[scale=0.4]{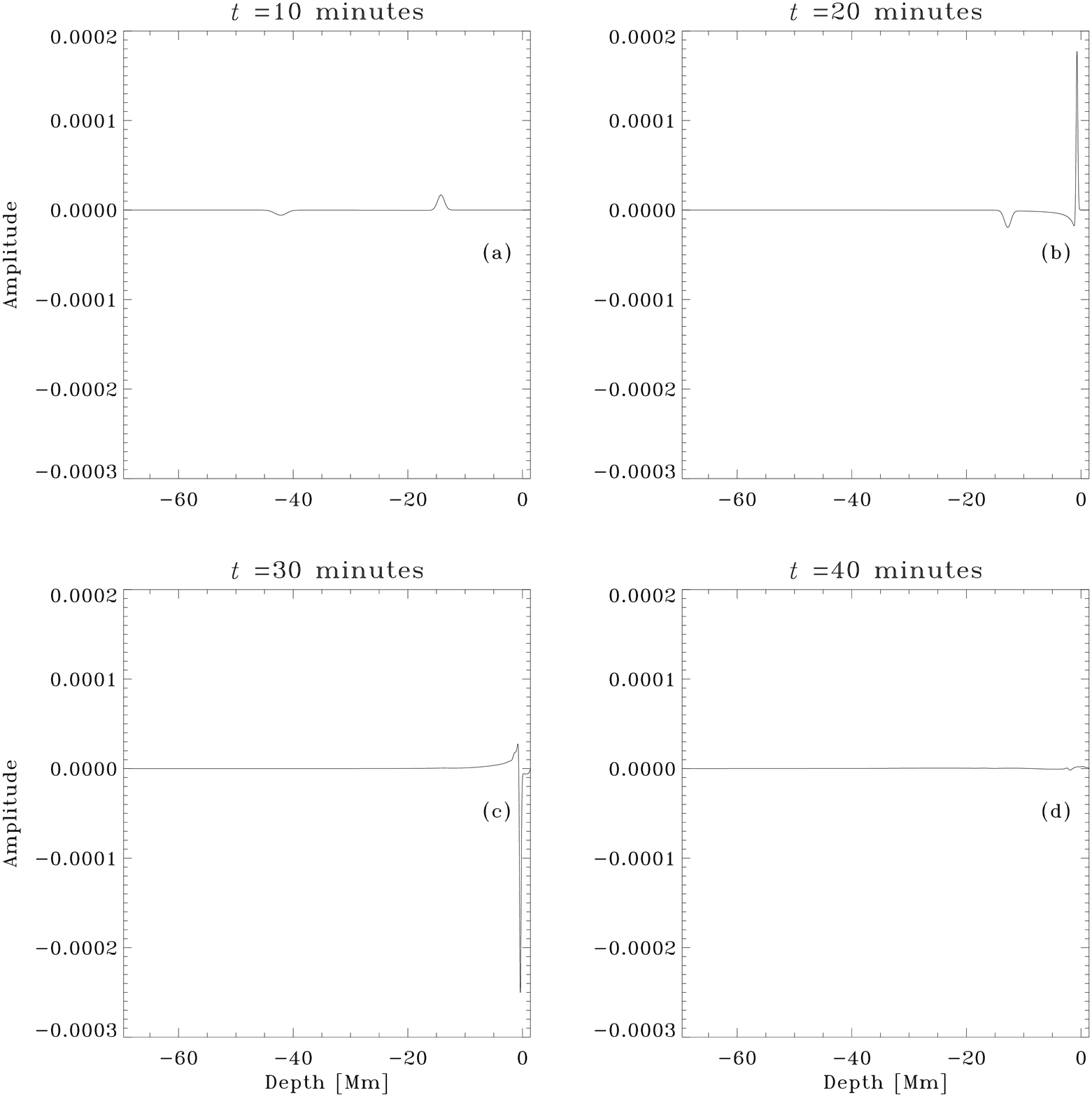}
\end{center}
\caption{Efficacy of the transmitting boundary. The initial condition is a Gaussian-shaped velocity impulse. Panel a shows the situation at $t = 10$ min, and the successive panels show the impulses at later instants in time. The amplitude in panel d is of the order of $10^{-6}$, significantly smaller than in panels a through c. Together with the test of Figure~\ref{test.eig.bc}, the boundary seems to do a relatively good job of removing outward propagating waves while the interior portion of the eigenfunction is seen to be mostly undisturbed.}\label{working_bcs}
\end{figure}

A rough test of the efficacy of the boundary conditions is shown in Figure~\ref{working_bcs}, where an initial Gaussian-shaped velocity impulse is allowed to propagate outward. Panel a shows the situation at $t = 10$ min, and the successive panels show the impulses at later instants in time. The amplitude in panel d is of the order of $10^{-6}$, significantly smaller than in panels a through c. Together with the test of Figure~\ref{test.eig.bc}, the boundary condition seems to allow outward propagating waves to exit the computational domain while leaving the eigenfunctions relatively undisturbed. A check of this sort was applied to choose the sponge for the real simulations. Since the polytrope + isothermal stratification near the surface is very similar to the model used in the simulations, and since the sponges are quite similar in structure, we expect that the eigenfunctions in the simulations are also well retrieved while the sponge damps the outward propagating waves.

\acknowledgements
S. M. Hanasoge and S. Couvidat were supported by NASA grants HMI NAS5-02139 and MDI NNG05GH14G. S. P. Rajaguru was supported by a Living With a Star program grant NNG 05-GM85G. A.C. Birch acknowledges support from NASA contract NNH04CC05C. The computations were performed on the Columbia supercomputer at NASA Ames. We thank Thomas Duvall, Jr. for useful discussions and the referee for suggestions that helped improve this paper.

\end{document}